\documentclass[12pt]{iopart}

\usepackage{graphicx}
\usepackage{iopams}
\usepackage{subeqn}
\usepackage{cite}

\newcommand{\ket}[1]{\left|#1\right\rangle}

\newcommand{\mean}[1]{\left\langle #1 \right\rangle}

\renewcommand{\Im}{{\rm Im}\,}
\renewcommand{\Re}{{\rm Re}\,}
\renewcommand{\imath}{\mathrm{i}}
\renewcommand{\e}{\mathrm{e}}

\newcommand{\eps}{\varepsilon}
\newcommand{\tp}{\tau}
\newcommand{\VH}{V_{0}}
\newcommand{\VL}{V_{1}}
\newcommand{\Np}{N}
\newcommand{\Nmanip}{N_{1}}
\newcommand{\Nmeas}{N_{0}}
\newcommand{\Jstat}{J_0}
\newcommand{\So}{\mathrm{L}}
\newcommand{\Dr}{\mathrm{R}}
\newcommand{\Le}{\mathrm{l}}
\renewcommand{\Le}{\ell}
\newcommand{\Ri}{\mathrm{r}}
\newcommand{\GammaSo}{\Gamma_\mathrm{L}}
\newcommand{\GammaDr}{\Gamma_\mathrm{R}}

\newcommand{\DM}{\ensuremath{\varrho}}			% full density matrix
\newcommand{\RDM}{\ensuremath{\sigma}}			% reduced system density matrix

\newcommand{\COMM}[2]{\left[#1, #2 \right]_-}	% commutator
\newcommand{\eqref}[1]{(\ref{#1})}

\begin{document}
%%%%%%%%%%%%%%%%%%%%%%%%%%%%%%%%%%%%%%%%%%%%%%%%%%%%%%%%%%%%%%%%%%%%%%%%%%%%%%%
%%% TITLE
%%%%%%%%%%%%%%%%%%%%%%%%%%%%%%%%%%%%%%%%%%%%%%%%%%%%%%%%%%%%%%%%%%%%%%%%%%%%%%%
\title{Coherent manipulation of charge qubits\\ in double quantum dots}
\author{Alexander Croy and Ulf Saalmann}
\address{Max Planck Institute for the Physics of Complex Systems\\
           N\"{o}thnitzer Str.~38, 01187 Dresden, Germany}
\ead{croy@pks.mpg.de}

\begin{abstract}	
The coherent time evolution of electrons in double quantum dots induced by fast bias-voltage switches is studied theoretically.
As it was shown experimentally, such driven double quantum dots are potential devices for controlled manipulation of charge qubits.
By numerically solving a quantum master equation we obtain the energy- and time-resolved electron transfer 
through the device which resembles the measured data.
The observed oscillations are found to depend on the level offset of the two dots during the manipulation and, most surprisingly, also the on initialization stage.
By means of an analytical expression, obtained from a large-bias model, we can understand the prominent features of 
these oscillations seen in both the experimental data and the numerical results. These findings strengthen the common 
interpretation in terms of a coherent transfer of electrons between the dots.
  \end{abstract}
  % Quantum dots, electronic transport in, 73.63.Kv
  % Coulomb blockade, 73.23.Hk
  % Theory of electronic transport; scattering mechanisms 72.10.-d
  \submitto{\NJP}

  \maketitle
  \tableofcontents
  
%%%%%%%%%%%%%%%%%%%%%%%%%%%%%%%%%%%%%%%%%%%%%%%%%%%%%%%%%%%%%%%%%%%%%%%%%%%%%%%
%%% INTRODUCTION
%%%%%%%%%%%%%%%%%%%%%%%%%%%%%%%%%%%%%%%%%%%%%%%%%%%%%%%%%%%%%%%%%%%%%%%%%%%%%%%
\section{Introduction}\label{sec:Intro}
Coherent control of nanoscale devices is one of the main topics of
current research on electron transport in systems on the nanometer
scale. A manifest realization of control is given by pump-probe schemes known from molecular physics \cite{ze88,sm99}. In the context of electron transport the ``pump'' and ``probe'' steps consist in switching to and from a regime, where transport is largely
blocked \cite{fuha+06}. This provides an effective decoupling of the electronic system from the 
reservoirs and allows for a coherent evolution between pump and probe triggers. Such a 
setup has been successfully used to coherently control charge \cite{hafu+03} and 
spin qubits \cite{pejo+05,kobu+06} in double quantum dots (DQDs). 

The theoretical description of these experiments is a very demanding task, since a
fully time-resolved calculations are necessary. Although several formalisms exist to address this issue, only very few numerical schemes are available in this context. 
Typically, the numerical approaches to time-dependent electron
transport with arbitrary driving rely either on equations of motion for
non-equilibrium Green functions (NEGF) \cite{zhma+05,kust+05,hohe+06,mogu+07,stpe+08,crsa09a} or generalized quantum master
equations (QME) for the reduced density matrix \cite{wesc+06,jizh+08}.
Such calculations are very helpful to gain
a deeper understanding of the mechanisms relevant for coherent control, since
time-resolved quantities, such as occupations and currents, are readily accessible.

In this article we concentrate on the experiment of Fujisawa and coworkers on coherent
manipulation of charge states in double quantum dots \cite{hafu+03}. One of the hallmarks
of the experiment is the observation of oscillations of the so-called number of 
pulse-induced tunneling electrons as a function of pulse length (see also Sec.\,\ref{sec:ChQubits}). These oscillations are commonly associated with coherent tunneling
processes between the two quantum dots during the manipulation stage. While this picture
qualitatively explains the main features of the experiments, it neglects the influence of
the initialization and the measurement stages on the coherent time-evolution. As we will
show, consistently considering the whole pump-probe
sequence provides even stronger evidence for actual coherent control. Moreover,
the new insights may help to gain a deeper understanding of the dynamics
in nanoscale devices.

First of all we briefly introduce the concept of charge qubits in
double quantum dots and review the main results of the coherent manipulation
experiment.
The theoretical model and the relevant tools used in this work are
presented in Sec.\,\ref{sec:Theory}. Numerical results and a detailed analysis
using an analytically solvable model are given in Sec.\,\ref{sec:Results}. The 
article concludes with a summary in Sec.\,\ref{sec:concl}.

%%%%%%%%%%%%%%%%%%%%%%%%%%%%%%%%%%%%%%%%%%%%%%%%%%%%%%%%%%%%%%%%%%%%%%%%%%%%%%%
%%% CHARGE QUBITS
%%%%%%%%%%%%%%%%%%%%%%%%%%%%%%%%%%%%%%%%%%%%%%%%%%%%%%%%%%%%%%%%%%%%%%%%%%%%%%%
\subsection{Charge Qubits in Double Quantum Dots}\label{sec:ChQubits}

From quantum information theory it is well known that in principle any
two-level system may be used as a single bit of (quantum) information, which
is then called {\it qubit} \cite{nich00}. In laterally coupled quantum dots
one may find such two-level systems by considering charge states denoted by 
$(N_\Le, N_\Ri)$,
i.\,e., systems with $N_\Le$ electrons in the left and $N_\Ri$ electrons
in the right dot. An electron tunneling from left to right corresponds to the
sequence $(N_\Le, N_\Ri)\to (N_\Le{+}1, N_\Ri)
      \to (N_\Le, N_\Ri{+}1)\to (N_\Le, N_\Ri)$.
The probability for such a tunneling event is largest if the two charge states,
$\ket\Le= (N_\Le{+}1, N_\Ri)$ and 
$\ket\Ri=(N_\Le, N_\Ri{+}1)$
have a vanishing energy difference $\eps$ \cite{koma+97,wide+02}. In the
vicinity of such a resonance the coupled quantum dots can be described by a
two-level system, which is characterized by the energy difference $\eps$
and the interdot tunnel coupling $T_{\rm c}$ \cite{wide+02,hafu+03}. Correspondingly, the associated qubit is called {\it charge qubit}.

\begin{figure}[bt]
  \centering
  \includegraphics[width=.75\textwidth]{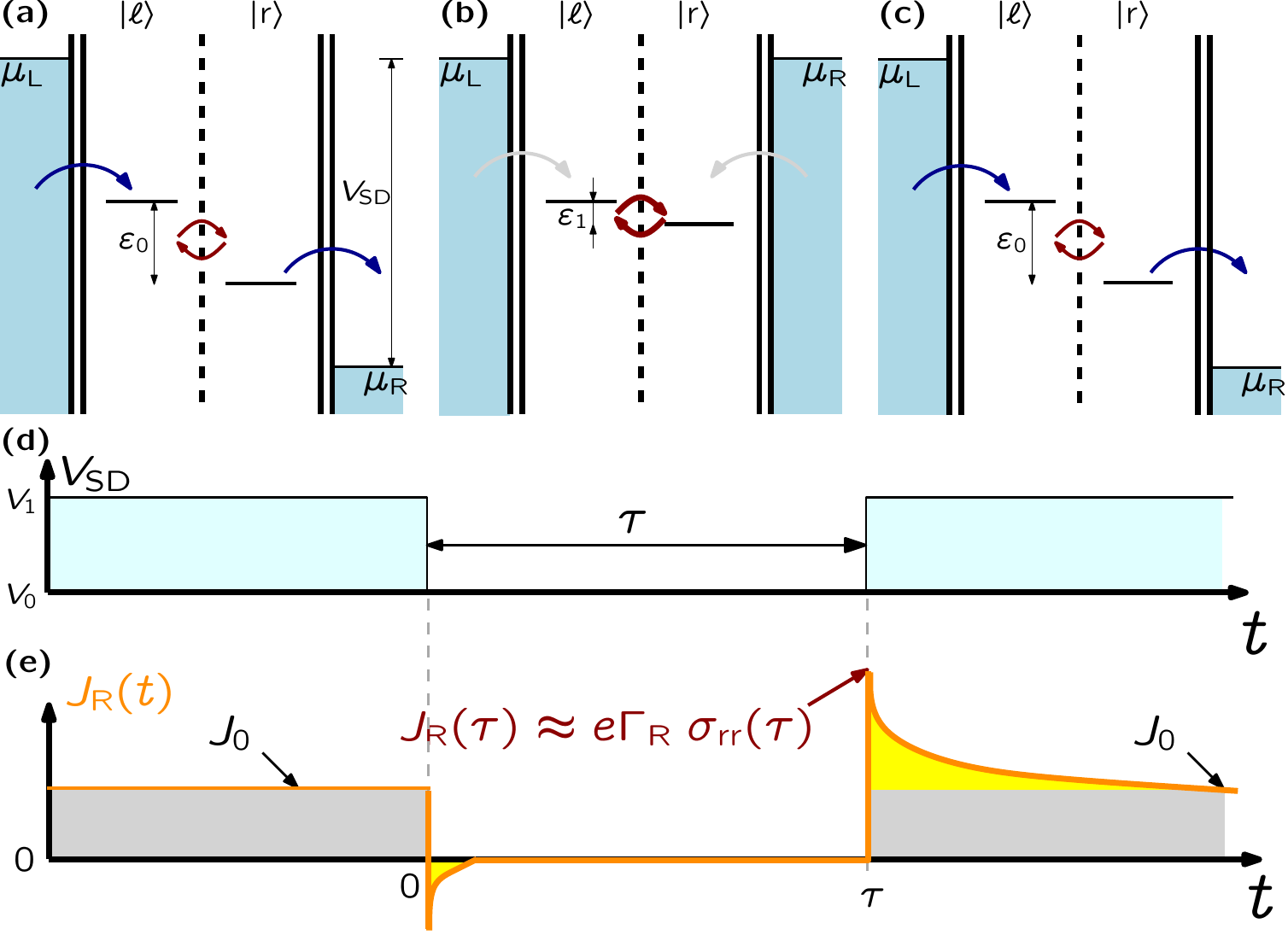}
  \caption{Pump-probe scheme for the coherent manipulation of a charge qubit 
  	\cite{hafu+03}.
  	{\bf a)} Initialisation ($V_{\rm SD}=\VH$), 
	{\bf b)} manipulation ($V_{\rm SD}=\VL$)
	and {\bf c)} measurement phase ($V_{\rm SD}=\VH$). 
	{\bf d)} Time dependence of the bias-voltage 
	pulse $V_{\rm SD}(t)$.
	{\bf e)} Current (orange line) as a function of time. The stationary value $\Jstat$ is taken before the manipulation phase, $t<0$, and for long times, $t\to\infty$. The number of transferred electrons $\Np$ (yellow-shaded area), without the stationary portion (gray-shaded),  has contributions from the manipulation and the measurement phase.
	}
  \label{fig:DQDPumpProbe}
\end{figure}

The Hamiltonian of the two coupled charge-states in the DQD, i.\,e.\ two coupled energy levels, reads 
\begin{equation}
	H_\mathrm{DQD} = \eps_\Le(t) \, c^\dagger_\Le c_\Le 
  + \eps_\Ri(t) \, c^\dagger_\Ri c_\Ri 
  + T_{\rm c} (c^\dagger_\Le c_\Ri + c^\dagger_\Ri c_\Le)
  + U c^\dagger_\Le c_\Le c^\dagger_\Ri c_\Ri\,.
	\label{eq:DQDHam}
\end{equation}
The operators $c^\dagger_{n}$ ($c_{n}$) create (annihilate) an electron in the
left ($n={\Le}$) or right ($n={\Ri}$) dot, respectively. The interdot
charging energy $U$ suppresses double occupancy of the DQD. The time-dependence of
the energies $\eps_{\Le}$ and $\eps_{\Ri}$ may be given by external gate-voltages. Figure \ref{fig:DQDPumpProbe}
shows the corresponding energy scheme for a charge qubit additionally coupled to 
a source and a drain reservoir.
In order to use a quantum system as a qubit, it has to fulfill at least the following 
three conditions \cite{di00}: (i) initialization of the qubit into a well defined state,
(ii) application of unitary operations ({\it quantum gates}) and (iii) readout
of the qubit state. Using a pump-probe scheme with rectangular bias-voltage pulses, these
three steps have successfully been implemented for a charge qubit in a double
quantum dot \cite{hafu+03}. The scheme reminds in many ways of the usual pump-probe 
experiments with atoms or molecules \cite{grze90,cokr07,kriv09}. In the present case the 
raising edge of the pulse triggers the coherent dynamics (pump), while the trailing edge
starts the measurement (probe). 

The main idea of the experiment consists in suddenly switching between the Coulomb blockade regime and a 
transport regime by using a bias-voltage pulse. In the former case the DQD is effectively 
isolated from the reservoirs since sequential tunneling is strongly suppressed \cite{grde92}. 
This provides the possibility to coherently control the charge qubit \cite{fuha+06}. The transport regime is used
to initialize the system and to readout the charge state after manipulation. The
whole sequence is schematically shown in Fig.\,\ref{fig:DQDPumpProbe}. 
Before and after the pulse, i.\,e.\ during the initialization and measurement phase, the source-drain voltage is $V_{\rm SD}=\VH$ and 
transport through the dot is possible since both charge states are within the transport 
window defined by the source-drain voltage. During the pulse, i.\,e.\ during the manipulation phase, the source-drain voltage is switched to $V_{\rm SD}=\VL$ and no transport is possible.

In the experiment, the time-averaged current is measured as a function of pulse length $\tp$ and
energy difference $\eps\equiv\eps_\Ri-\eps_\Le$. The latter is tuned by applying
a suitable gate-voltage $V_{\rm g}$ to the right quantum dot (not shown in Fig.\,\ref{fig:DQDPumpProbe}). 
Notice, that due to capacitive couplings \cite{hafu+03}, the energy difference $\eps$ for a given gate-voltage $V_{\rm g}$ 
also changes with the source-drain voltage, which is why we use $\eps_0$ and $\eps_1$ 
for high and low source-drain voltage, respectively, in Figs.\,\ref{fig:DQDPumpProbe}a--c.
The pulse is repeated with a repetition rate $f_{\rm rep}=100\,{\rm MHz}$ and by using a lock-in technique the pulse-induced current $J_{\rm p}$ is obtained, which does not contain the asymptotic (stationary)
current of the initialization and measurement phase. With $J_{\rm p}$ one finally
gets the {\it number of pulse-induced tunneling electrons} 
$\Np = J_{\rm p}/e f_{\rm rep}$ \cite{hafu+03}. It is then argued that this
quantity is equivalent to the occupation of the right dot at the end of the
pulse \cite{hafu+03}. As we will show in Sec.\,\ref{sec:model}, this assumption is not 
always valid. The experimental results show pronounced oscillations of
$\Np$ as a function of pulse length. Hereby, the frequency and the amplitude
of the oscillations depend on the energy difference between the charge states. These
oscillations are interpreted as a signature of coherent tunneling between the 
charge states (Rabi oscillations). Apart from the oscillations there are two other 
noteworthy features in the experimental results: for a fixed pulse length the function 
$\Np$ is asymmetric around $\eps_1 = 0$ and, in particular 
for $\eps_0=0$, the number of pulse-induced electrons takes negative values.
It has been argued that these features are artifacts of incomplete initialization
or imperfect pulse shapes \cite{hafu+03}. In the remaining part of the
article we will demonstrate, that this must not be the case and the additional features 
instead strengthen the view of a coherent manipulation.

%%%%%%%%%%%%%%%%%%%%%%%%%%%%%%%%%%%%%%%%%%%%%%%%%%%%%%%%%%%%%%%%%%%%%%%%%%%%%%%
%%% THEORETICAL DESCRIPTION
%%%%%%%%%%%%%%%%%%%%%%%%%%%%%%%%%%%%%%%%%%%%%%%%%%%%%%%%%%%%%%%%%%%%%%%%%%%%%%%
\section{Theoretical Description}\label{sec:Theory}

After specifying the Hamiltonian we briefly present the time-dependent description of the DQD 
using a quantum master equation. Furthermore we discuss a model which allows for deriving analytical expressions 
and facilitates a detailed analysis. 
Specifically, we consider the large-bias limit and use a description of the dynamics, which has originally been
developed for photon-assisted transport through DQDs \cite{stna96}.

\subsection{Setup}
Apart from the DQD Hamiltonian [Eq.\,\eqref{eq:DQDHam}], the total Hamiltonian also
contains terms describing the electron reservoirs and the tunnel coupling,
\begin{equation}
	H = H_\mathrm{DQD} + H_\mathrm{res} + H_\mathrm{tun}\,.
	\label{eq:QDStotalHam}
\end{equation}
The reservoirs are described as usual by non-interacting electrons,
\begin{equation}
  H_\mathrm{res} = \sum_{\alpha = \So,\Dr} 
  \sum_k \eps_{\alpha k}(t) b^\dagger_{\alpha k} b_{\alpha k}\;,
  \label{eq:QDSResHamilOp}
\end{equation}
where $\{b^\dagger_{\alpha k}\}$ and $\{b_{\alpha k}\}$ are the electron creation and annihilation operators for the $\alpha$-reservoir state $k$, respectively.   
The reservoir single-particle energies have the general form $\eps_{\alpha k}(t) = \eps^0_{\alpha k} + \Delta_{\alpha}(t)$ with the $\Delta_{\alpha}$ accounting for a time-dependent bias.
The tunneling Hamiltonian for the linear setup of two quantum dots in series, which are
each coupled to a single reservoir, reads explicitly
\begin{eqnarray}\label{eq:QDSTunnHam}
  H_\mathrm{tun} &=&
  	\sum_k \left[
	T_{\So k, {\Le}} b^\dagger_{\So k} c_\Le +  
	T_{\Dr k, {\Ri}} b^\dagger_{\Dr k} c_\Ri
	+ \rm{h.c.}
	\right] \\
	&=& \sum_{\alpha=\So,\Dr} \sum_{n=\Le,\Ri}
		\left[ 
		B^{(+)}_{\alpha n} S^{(+)}_n + S^{(-)}_n B^{(-)}_{\alpha n}
		\right]\;,
\end{eqnarray}
with $T_{\alpha k, n}$ denoting the coupling matrix element between QD $n=\Le, \Ri$ and the $k$-th
mode of the respective reservoir $\alpha={\rm \So,\Dr}$. In the second line of the
tunneling Hamiltonian we have introduced the abbreviations 
$B^{(+)}_{\alpha n} =\sum_k T_{\alpha k, n} b^\dagger_{\alpha k}$,
$B^{(-)}_{\alpha n} =\sum_k T^*_{\alpha k, n} b_{\alpha k}$ and
$S^{(+)}_n = c_n$, $S^{(-)}_n = c^\dagger_n$, which will simplify the notation later on.

Further, making the wide-band assumption
renders the tunnel-coupling elements independent of the reservoir state, $T_{\alpha k, n} = T_{\alpha, n}$, which yields
a flat spectral density,
\begin{equation}
  \Gamma_{\alpha m n} = 
  2 \pi T_{\alpha, n} T^*_{\alpha, m} \sum_k
  \delta(\eps - \eps^0_{\alpha k})\;.
  \label{eq:QDSLevelWidth}
\end{equation}

\subsection{Time-local Quantum Master Equation (TLQME)}\label{sec:TLQME}
The time-evolution of the density operator $\DM$ characterizing the total system
is determined by the Liouville-von\,Neumann equation,
\begin{equation}\label{eq:OSLvNEq}
  \imath \hbar \frac{\partial }{\partial t} \DM = \COMM{ H }{ \DM }\;,
\end{equation}
where $H$ is given by Eq.\,\eqref{eq:QDStotalHam}. Since we are only interested in the 
properties of the DQD, we can trace out the reservoir degrees of freedom and thus obtain 
the reduced density operator $\RDM = \Tr_{\rm res} \DM$. In order to get an equation
of motion for $\RDM$, we employ the {\it time-convolutionless projection operator 
technique} \cite{chsh79,shar80,brpe02}, which yields a time-local differential equation 
for $\RDM$ \cite{wesc+06},
\begin{eqnarray}\label{eq:QMETCL}
\imath \frac{\partial}{\partial t}\RDM(t)
&=&\COMM{H_{\rm S}}{\RDM(t)} -\imath \sum\limits_\alpha\sum\limits_{m}\left(
    \COMM{S^{(+)}_{m}}{ \Lambda^{(+)}_{\alpha m}(t) \, \RDM(t)
      - \RDM(t) \, \tilde{\Lambda}^{(+)}_{\alpha m}(t)} \right. \nonumber\\
&&\left.   +\COMM{S^{(-)}_{m}}{\Lambda^{(-)}_{\alpha m}(t) \, \RDM(t)
   - \RDM(t)\, \tilde{\Lambda}^{(-)}_{\alpha m}(t)}
  \right)\;.
\end{eqnarray}
Here and in the following we set $\hbar =1$.
The TLQME is supplemented by the following auxiliary operators \cite{wesc+06}
\begin{subequations}\label{eq:QMEDefLAuxOp}
\begin{eqnarray}
  \Lambda^{(x)}_{\alpha m} (t) &=& \sum\limits_{y=+,-} \;
  \int\limits_{t_0}^{t} dt' C_{\alpha m}^{(x y)}(t,t') \;U^\dagger_\mathrm{S}(t,t') S_m^{(y)} U_\mathrm{S}(t,t') \;,\\
  \tilde{\Lambda}^{(x)}_{\alpha m}(t) &=& \sum\limits_{y=+,-} \;
  \int\limits_{t_0}^{t} dt' C_{\alpha m}^{(x y)*}(t,t') \;U^\dagger_\mathrm{S}(t,t') S_m^{(y)} U_\mathrm{S}(t,t')\;.
\end{eqnarray}
\end{subequations}
These auxiliary operators are determined by the free DQD propagator $U_\mathrm{S}(t,t')$
and the reservoir correlation functions $C_{\alpha m}^{(x y)}(t,t')$. The former
is given in terms of the DQD Hamiltonian, 
$U_\mathrm{S}(t,t') = \mathcal{T} \exp\left(-\imath \int^t_{t'} dt'' H_{\rm DQD}(t'') \right)$, where
$\mathcal{T}$ is the usual time-ordering prescription.
The correlation functions
describe the influence of the reservoir on the system dynamics. In the current context 
they are given by  \cite{wesc+06}
\begin{subequations}\label{eq:QMEDefCorrFunc}
\begin{eqnarray}
  C_{\alpha m}^{(+ -)}(t,t') &=& 
  		\int \frac{d\eps}{2\pi} \Gamma_{\alpha m m} f_{\alpha}(\eps) 
			\exp\left( \imath \int^t_{t'} dt'' \left[\eps {+} \Delta_{\alpha}(t'') \right]\right)\;,\\
  C_{\alpha m}^{(- +)}(t,t') &=& \int \frac{d\eps}{2\pi} \Gamma_{\alpha m m}  
  		\left[1 {-} f_{\alpha}(\eps) \right]
			\exp\left( -\imath \int^t_{t'} dt'' \left[\eps {+} \Delta_{\alpha}(t'') \right] \right)
  \;.
\end{eqnarray}
\end{subequations}
The Fermi function $f_{\alpha}(\eps)$ characterizes the initial 
state of reservoir $\alpha$, and the spectral density $\Gamma_{\alpha m m}$ has been defined in Eq.\,\eqref{eq:QDSLevelWidth}. In order to include the energy-level broadening induced by
the back-action of the reservoirs on the DQD, we modify the correlation functions by multiplying
with an exponential decay factor \cite{brsc94,ovne04,lilu+05}. Consequently,
we have $C_{\alpha m}^{(x y)}(t, t') \propto \exp\left( -\Gamma_{\alpha m m} (t{-}t')/2 \right)$
\cite{ovne04}.

Although the equation of motion for the reduced density matrix is local in time,
it is still very demanding to solve it numerically. An efficient 
method to propagate $\RDM$ is the so-called {\it auxiliary-mode expansion technique},
which is based on a decomposition of the Fermi function \cite{wesc+06}. As a consequence the
operators $\Lambda^{(x)}$ and $\tilde{\Lambda}^{(x)}$ are also decomposed into a sum
and for each term an individual equation of motion can be derived. The details
of this procedure are given in \ref{sec:AuxModeProp}. The auxiliary-mode expansion
has successfully been used in the context of time-nonlocal and time-local
quantum master equations \cite{meta99,wesc+06,jizh+08} and has recently been applied to
non-equilibrium Green functions \cite{crsa09a}.

It remains to discuss the calculation for the time-resolved electric current 
through the tunneling barriers in a way consistent with the QME \cite{leko+02}.
The electric current through the tunneling barrier $\alpha$ is given by the rate 
of change of the particle number in reservoir $\alpha$ \cite{wija+93,brsc94}, 
\begin{eqnarray}
  J_\alpha (t) 
  &=& -e \frac{d}{dt}\mean{N_\alpha} 
  = -\imath e \mean{ \left[ H, N_\alpha \right] } \nonumber\\
  &=& -2 e \,\Im 
  		\left[ \Tr \left\{ 
  			\sum\limits_m B^{(+)}_{\alpha m} S^{(+)}_m \DM(t) 
		\right\}\right]\;.
  \label{eq:CurrDef}		    	
\end{eqnarray}
Plugging the formal solution of the Liouville-von\,Neumann equation [Eq.\,\eqref{eq:OSLvNEq}] in the
interaction representation into Eq.\,\eqref{eq:CurrDef} and keeping only terms up to 2nd 
order in $H_\mathrm{tun}$, one finds within the time-local approximation for the 
time-dependent current \cite{wesc+06},
\begin{eqnarray}\label{eq:TDCurrent}
  J_\alpha (t) = 2 e\, \Re \sum\limits_{m} \Tr_\mathrm{DQD} 
  	\left\{
    	c_m \Lambda^{(+)}_{\alpha m}(t) \RDM(t)
    - c_m \RDM(t) \tilde{\Lambda}^{(+)}_{\alpha m}(t) 
   \right\} \,.
\end{eqnarray}
This expression is consistent with the TLQME, which is also of second order in $H_\mathrm{tun}$.
Note, that in order to calculate the current one only needs the auxiliary operators 
$\Lambda^{(+)}_{\alpha m}$, $\tilde{\Lambda}^{(+)}_{\alpha m}$ and the reduced density operator $\RDM$.

\subsection{Markovian Approximation in the Large-Bias Limit (LBL)}\label{sec:Markovian}
Considering the experimental situation, one is lead to an even simpler description of 
the dynamics in the DQD. Firstly, for a large inter-dot interaction strength $U$
only the following three states are relevant: $\ket{0}, \ket{\Le} = c^\dagger_\Le\ket{0}$ and $\ket{\Ri}= c^\dagger_\Ri\ket{0}$ \cite{stna96}. These correspond to an empty DQD, one electron occupying the left dot and
one electron occupying the right dot, respectively. These three states are used
as a basis for the reduced density operator in the following considerations. 
The second simplification arises due to the large bias in the experiment. In this case, using the Markovian limit of the QME [Eq.\,\eqref{eq:QMETCL}] is well justified \cite{stna96,gupr96}.

For the initialization and measurement phase (cf.\ Fig.\,\ref{fig:DQDPumpProbe}) one obtains the following differential equations for the components of the reduced density matrix \cite{na93,stna96}
\begin{subequations}\label{eq:DQDNazarovDynEq}
\begin{eqnarray}
  \dot{\RDM}_{\Le\Le} &=& -\imath T_{\rm{c}} \left( \RDM_{\Le\Ri} {-}
    \RDM_{\Ri\Le}\right) + \GammaSo \left( 1 {-} \RDM_{\Le\Le} {-} \RDM_{\Ri\Ri}\right)\;, \\
  \dot{\RDM}_{\Ri\Ri} &=& -\imath T_{\rm{c}} \left( \RDM_{\Ri\Le} {-}  
    \RDM_{\Le\Ri}\right) - \GammaDr \RDM_{\Ri\Ri}\;, 	\\
  \dot{\RDM}_{\Le\Ri} &=& -\imath T_{\rm{c}} \left( \RDM_{\Le\Le} {-} 
    \RDM_{\Ri\Ri}\right) +\imath \eps_0 \RDM_{\Le\Ri} 
  - (\GammaDr/2) \RDM_{\Le\Ri}\;.
\end{eqnarray}
\end{subequations}
Obviously, the coupling to the reservoirs introduces transitions between the
charge states $\ket{0}$, $\ket{\Le}$ and $\ket{\Ri}$. The first two equations 
describe the evolution of the charge-state occupations, which change due to tunneling
between the two dots and because of tunneling from and to the reservoirs. The remaining
equation yields the dynamics of the coherences $\RDM_{\Ri\Le} = \RDM_{\Le\Ri}^*$. 
Tunneling out of the DQD leads to loss of coherence.

In the manipulation phase an effective decoupling of the DQD from the reservoirs
is achieved by switching the source-drain voltage. An electron can still enter the
DQD, but leaving the system is strongly suppressed. In contrast to the initialization
phase, where tunneling out of the DQD leads to a vanishing coherence, here the dynamics
stays approximately coherent. Therefore, we assume the following equations of motion
during the voltage pulse,
\begin{subequations}\label{eq:BlochEq}
\begin{eqnarray}
  \dot{\RDM}_{\Le\Le} &=& -\imath T_{\rm{c}} \left( \RDM_{\Le\Ri} {-}
    \RDM_{\Ri\Le}\right) - \gamma \RDM_{\Le\Le} + \GammaSo \left(
    1 {-} \RDM_{\Le\Le} {-} \RDM_{\Ri\Ri}\right)\;, \\
  \dot{\RDM}_{\Ri\Ri} &=& -\imath T_{\rm{c}} \left( \RDM_{\Ri\Le} {-}  
    \RDM_{\Le\Ri}\right) - \gamma \RDM_{\Ri\Ri} + \GammaDr \left(
    1 {-} \RDM_{\Le\Le} {-} \RDM_{\Ri\Ri}\right)\;, \\
  \dot{\RDM}_{\Le\Ri} &=& -\imath T_{\rm{c}} \left( \RDM_{\Le\Le} {-} 
    \RDM_{\Ri\Ri}\right) +\imath \eps_1 \RDM_{\Le\Ri} - \gamma \RDM_{\Le\Ri}\;.
\end{eqnarray}
\end{subequations}
The rate $\gamma$ has been introduced to account for additional processes leading to
decoherence, such as background-charge fluctuations or back-action of the reservoirs
on the DQD. In principle, these processes are also present in the other stages, but
there they are dominated by the transport from source to drain.
The expressions for the time-resolved currents are very simple and
may, for instance, be extracted from Eqs.\ \eqref{eq:DQDNazarovDynEq} and \eqref{eq:BlochEq}.
They are explicitly given by
\begin{subequations}
\begin{equation}
  J_\Dr (t) = e\,\GammaDr \RDM_{\Ri\Ri} (t)\;,
\end{equation}\begin{equation}
  J_\Dr (t)
  = - e\,\GammaDr \left[ 1  {-} \RDM_{\Le\Le} (t) {-} \RDM_{\Ri\Ri} (t) \right] + e\,\gamma\,\RDM_{\Ri\Ri} (t) \;,
\end{equation}
\end{subequations}
in the initialization/measurement and manipulation phase, respectively 
\cite{stna96,gupr96}.

%%%%%%%%%%%%%%%%%%%%%%%%%%%%%%%%%%%%%%%%%%%%%%%%%%%%%%%%%%%%%%%%%%%%%%%%%%%%%%%
%%% RESULTS & DISCUSSION
%%%%%%%%%%%%%%%%%%%%%%%%%%%%%%%%%%%%%%%%%%%%%%%%%%%%%%%%%%%%%%%%%%%%%%%%%%%%%%%
\section{Results and Discussion}\label{sec:Results}
In the following we will discuss a DQD system with parameters based on the experimental values \cite{hafu+03}. 
We summarize all relevant quantities in Table\,1.
As sketched in Fig.\,\ref{fig:DQDPumpProbe}, we consider a perfect rectangular pulse with
duration $\tp$ and assume that the DQD is an stationary state at $t=0$, when the manipulation phase starts.
In the numerical calculation, at $t=0$ the source-drain voltage
switches from $\VH$ to $\VL$ and at $t=\tp$ it switches back. The level of the left dot is fixed at $\eps_\Le = -e\VH/2$.
In order to get an energy-resolved picture we can shift the right level $\eps_\Ri = -\VH/2 + eV_{\rm g}$ by changing the gate voltage $V_{\rm g}$ at the right dot, similar to the experiment \cite{hafu+03}.
Due to capacitive couplings there is an additional offset $\delta\eps$ during the pulse and the energy of the right 
dot is $\eps_\Ri = -\VH/2  + eV_{\rm g} + \delta\eps$.
Thus, the level offset $\eps=\eps_\Ri-\eps_\Le$ is $\eps_{0}=eV_{\rm g}$ and $\eps_{1}=eV_{\rm g}+\delta\eps$.
In principle it is sufficient to specify either $\eps_{0}$ or $\eps_{1}$, since this fixes automatically the other one. 
However, for convenience we will use both notations in the following.   
\begin{table}[t]
\def\eq{\hspace*{-2em}$=$\hspace*{-2em}}
	\caption
    {Parameter values used for the numerical calculations,
     extracted from the experiment \cite{hafu+03}.}
    \label{tab:APPTranEnParams}
  \centering
 \begin{tabular*}{0.6\textwidth}[b]{@{\extracolsep{\fill}}lr@{}c@{}l}
  	\br
    \bf parameter & \multicolumn{3}{c}{\bf value} \\
    \mr
    capacitive level offset & $\delta\eps$& \eq & $30\,{\rm \mu eV}$\\
    interdot tunnel-coupling    &$T_{\rm c}$& \eq & $4.5\,{\rm \mu eV}$\\
%    Ladungsenergie      &$E_{\rm c}$& $ =$& $ 1.3\,{\rm m eV}$\\
	\mr
    tunnel rates      &$\GammaSo = \GammaDr$& \eq & $ 30\,{\rm \mu eV}$\\
    source-drain voltages &$\VH$& \eq& $ 650\,{\rm \mu V}$\\
                               &$\VL$& \eq & $ 0\,{\rm \mu V}$\\
    electron temperature & $1/\beta$ & \eq & $ 10\,{\rm \mu eV}$\\
    \mr
  \end{tabular*}
\end{table}

\subsection{Numerical Solution of the TLQME}
We have investigated the pump-probe scheme presented in Sec.\ \ref{sec:ChQubits}
using the TLQME given by Eqs.\ \eqref{eq:QMETCL} and \eqref{eq:QMEDefLAuxOp}. The operators are represented in the basis $\{\ket{0}, \ket{\Le},\ket{\Ri}\}$. The resulting system of differential equations is propagated by means of an auxiliary-mode expansion described in \ref{sec:AuxModeProp}.
In order to calculate the number of pulse-induced tunneling electrons the numerically
determined current [Eq.\,\eqref{eq:TDCurrent}] was integrated over time and the stationary contribution was subtracted.
Therefore, the following integral has to be calculated
\begin{equation}\label{eq:DQDNpTp}
  \Np (\eps,\tp) =  \frac{1}{e}\int\limits^{+\infty}_{-\infty } d t \left[ J_\Dr (\eps,t) - \Jstat(\eps) \right]
   + \frac{\Jstat(\eps)}{e} \tp\;.
\end{equation}
The dependence on $\eps$ accounts for both situations $\eps=\eps_{0}$ and $\eps=\eps_{1}$,
which have a fixed relation $\eps_{1}=\eps_{0}+\delta\eps$ as explained above.
$\Jstat$ refers to the stationary current reached at the end of the initialization and the measurement phase.
The last term has been added to correct for the different stationary current during the manipulation phase (cf.\ Fig.\,\ref{fig:DQDPumpProbe}e).

\begin{figure}[bt]
  \centering
  \includegraphics[width=.9\textwidth]{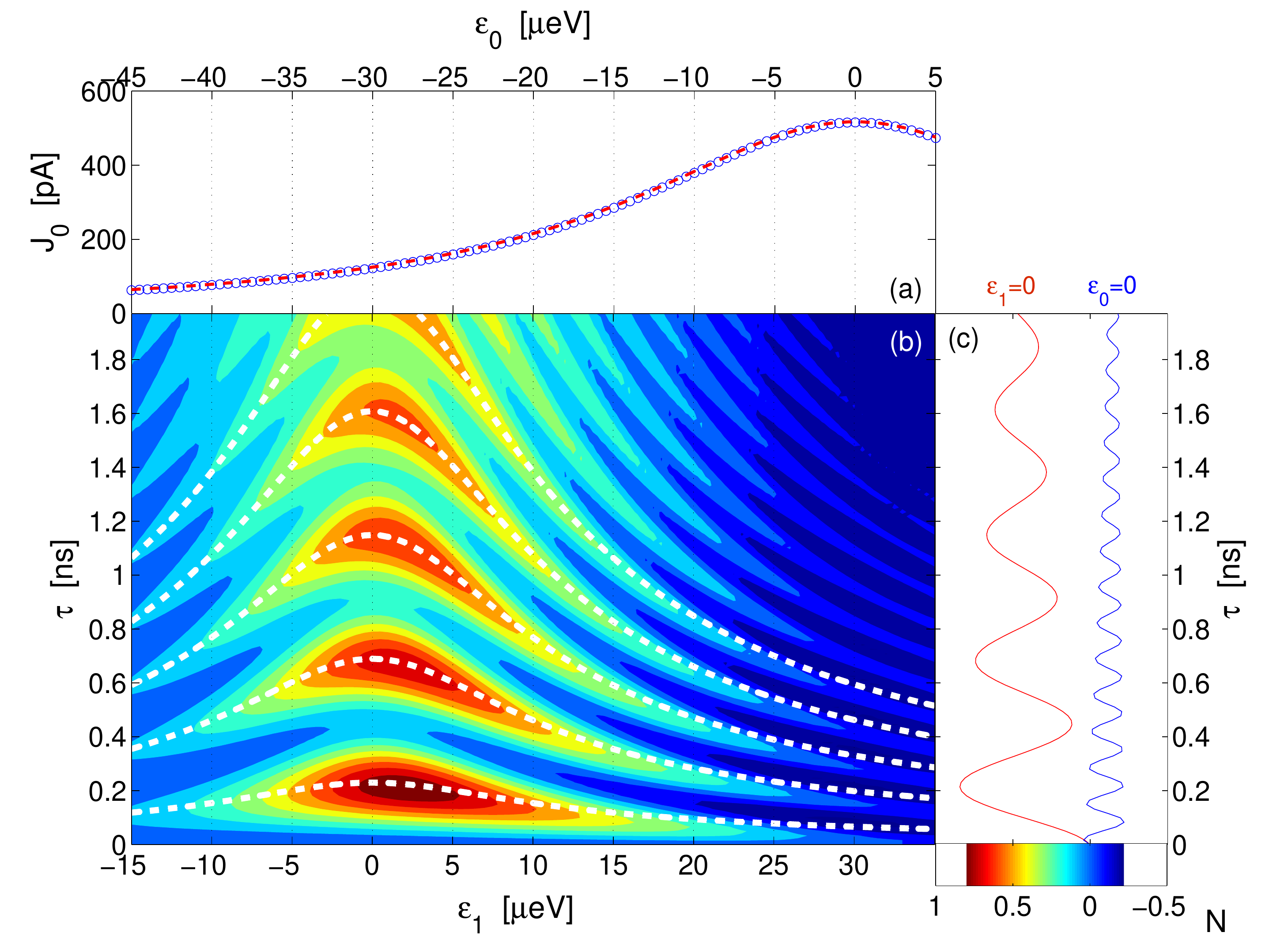}
  \caption{Numerical results for $\Jstat$ and $\Np$ as 
    functions of energy difference $\eps_{0/1}$ and pulse length $\tp$. 
    {\bf a)} Stationary current $\Jstat (\eps_{0}) $ before pulse.
    	 Symbols indicate numerical expressions and dashed lines denote analytical
	 	 expressions for the LBL, i.\,e.\ $V_{\rm SD}\to \infty$ \cite{na93,vago+95}.
    {\bf b)} Number of pulse-induced tunneling electrons $\Np$ calculated with
       Eq.\ \eqref{eq:DQDNpTp}. The dashed white lines show
       expected positions of maxima
       $\tp_{n}=(2 n {+} 1)\pi/\Omega$ with $n=0,1,\ldots$
    {\bf c)} Number of pulse-induced tunneling electrons $\Np$
       vs pulse length for constant energy difference $\eps_0=0$ und $
       \eps_1=0$.}
  \label{fig:DQDNumPopulationNp}
\end{figure}

Figure \ref{fig:DQDNumPopulationNp}a shows the current $\Jstat$ at the end of the
initialization phase as a function of the level offset $\eps_{0}$. This current is maximal when the energies of the two charge states
coincide ($\eps_0 = 0$). Also shown is the analytical result for large 
source-drain voltages \cite{na93,vago+95}.
In Fig.\,\ref{fig:DQDNumPopulationNp}b the number of pulse-induced electrons
$\Np$ is shown as a function of pulse length $\tp$ and energy difference
$\eps_{0/1}$. Hereby, the function $\Np (\eps_1,\tp)$ is asymmetric around $\eps_1=0$. 
Moreover, one can clearly observe an oscillatory behavior of $\Np$. The
frequency of the oscillations increases with increasing values of $\eps_1$. This
is explicitly shown in Fig.\,\ref{fig:DQDNumPopulationNp}c, where the function
$\Np (\eps,\tp)$ is shown for two energy differences, $\eps_0=0$ and 
$\eps_1=0$, respectively. For the latter case, the oscillations have the
largest amplitude.

Following the hypothesis that the number of tunneling electrons reflects just the 
occupation of the right dot at the end of the voltage pulse, one may naturally interpret 
the oscillations seen in Fig.\,\ref{fig:DQDNumPopulationNp}b as {\it Rabi oscillations}
between the two charge states \cite{hafu+03}, with the frequency given by 
\begin{equation}\label{eq:rabifreq}
\Omega=\sqrt{\eps_1{}^{2} + 4 T_{\rm c}{}^{2}}\,.
\end{equation}
Consequently, the maxima of 
$\Np$ should appear at pulse lengths $\tp_{n}=(2 n {+} 1)\pi/\Omega$ with $n=0,1,\ldots$,
whereby the frequency $\Omega$ depends on $\eps_{1}$ according to Eq.\,\eref{eq:rabifreq}. These positions are indicated by white dashed lines
in Fig.\,\ref{fig:DQDNumPopulationNp}b. Around $\eps_1 = 0$ the maxima of
$\Np$ are indeed found at the expected positions. However, for large
$\eps_1$ the maxima are clearly shifted compared to $\tp_{n}$.

\begin{figure}[bt]
  \centering
  \includegraphics[width=.7\textwidth]{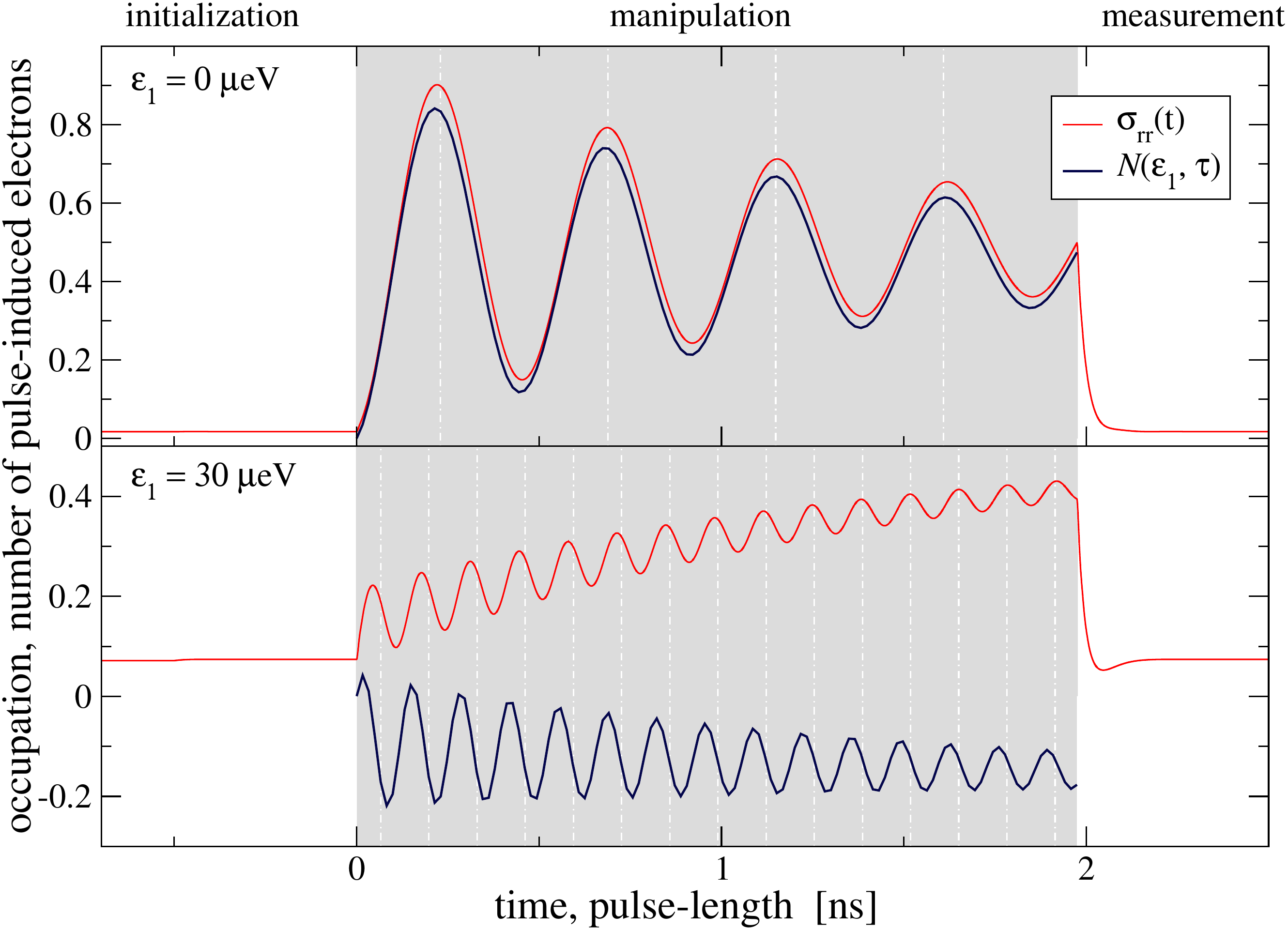}
  \caption{Occupation of the right quantum dot $\RDM_{\Ri\Ri}$ (red lines)
  	and the number of pulse-induced electrons $\Np$ (blue lines) as functions of time $t$ or pulse length $\tp$, respectively, from solving the TLQME numerically.  Maxima are expected at $t_{n}=\tp_{n}=(2 n {+} 1)\pi/\Omega$ with $n=0,1,\ldots$ (vertical dashed lines).}
  \label{fig:DQDNumPopulationNp2}
\end{figure}

It is interesting to compare the numerical data for $\Np (\eps,\tp)$ with the instantaneous occupation of the right dot $\RDM_{\Ri\Ri}(t)$. 
This is shown in Fig.\,\ref{fig:DQDNumPopulationNp2} for two energies $\eps$. For $\eps_1\,{=}\,0$ ($\eps_0\,{=}\,{-}30\,{\rm \mu eV}$)
both quantities are almost identical to each other. In the case $\eps_1\,{=}\,30\,{\rm \mu eV}$ ($\eps_0\,{=}\,0$) one observes considerable deviations between $\Np$ and $\RDM_{\Ri\Ri}$.
In contrast to the transferred electrons $N$, the occupation $\RDM_{\Ri\Ri}$ does have maxima at the times expected from the Rabi oscillations. 
Furthermore, $N$ also takes negative values.
Altogether, this is in clear contradiction to the assumption that the number of transferred electrons is just the occupation of the right dot.

In summary, the numerical results and the experimental observations are qualitatively very similar. In particular, the described features of the behavior of $\Np$ can also
be seen in the experimental results. However, the reasons for the peculiar features, 
the asymmetry of $\Np$  around $\eps_1=0$, the negative values of $\Np$
for large $\eps_1$ and the apparent shift of the maxima, cannot be explained by the numerical investigations alone. Therefore, in the remaining part of this section we 
will consider an analytically solvable model based on the Markovian quantum-master equations of Sec.\,\ref{sec:Markovian}.

\subsection{Model Calculations in the LBL}\label{sec:model}
Starting point of the analysis is the initialization phase described by 
Eq.\,\eqref{eq:DQDNazarovDynEq}. Its stationary state provides the input for
the manipulation phase. The stationary solution of Eq.\,\eqref{eq:DQDNazarovDynEq} 
directly gives the the initial state within the considered pump-probe scheme. 
Setting $\dot{\RDM}=0$ yields the stationary populations $\RDM_{\Le\Le}$, $\RDM_{\Ri\Ri}$ and coherences $\RDM_{\Le\Ri}$
\begin{subequations}\label{eq:DQDInitValues}
\begin{eqnarray}
    \RDM_{\Le\Le}(0) &=& 1 - \frac{8 T^2_{\rm{c}}}{12 T^2_{\rm{c}} + \Gamma^2 +
    4 \eps^2_0}\;, \\
  \RDM_{\Ri\Ri}(0) &=& \frac{4 T^2_{\rm{c}}}{12 T^2_{\rm{c}} + \Gamma^2 +
    4 \eps^2_0}\;, \\
  \RDM_{\Le\Ri}(0) &=& -\frac{2 \imath T_{\rm{c}} (\Gamma + 2 \imath \eps_0)}{12
    T^2_{\rm{c}} + \Gamma^2 + 4 \eps^2_0}\,.
\end{eqnarray}
\end{subequations}
The stationary state is in general a mixed state. Considering the case
of strong coupling to the reservoirs, $\Gamma \gg T_{\rm c}$, gives
\begin{subequations}\label{eq:DQDApprox}
\begin{eqnarray}
    \RDM_{\Le\Le}(0) &=& 1 - {\cal O}(T_{\rm c}^2)\;, \\
    \RDM_{\Ri\Ri}(0) &=& {\cal O}(T_{\rm c}^2)\;,\\
    \RDM_{\Le\Ri}(0)  &=& -\frac{2\imath T_{\rm c}}{\GammaDr-2\imath \eps_0} + {\cal O}(T_{\rm c}^2)\;.
\end{eqnarray}
\end{subequations}
Obviously, in this case the initialization leads to an almost perfect localization of an electron
in the left quantum dot. However, it is important to notice that at the same time
\emph{finite coherences} are unavoidable, which will have consequences for the manipulation phase.

For the manipulation phase it is convenient to introduce the following combinations of matrix elements of the density matrix $\RDM$,
\begin{equation}\begin{array}[b]{rclcrcl}
  s(t) &\equiv& \RDM_{\Le\Le} + \RDM_{\Ri\Ri}\;, &\quad&
  w(t) &\equiv& \RDM_{\Le\Le} - \RDM_{\Ri\Ri}\;, \\
  u(t) &\equiv& \RDM_{\Le\Ri} + \RDM_{\Ri\Le}\;, &\quad&
  v(t) &\equiv& -\imath \left( \RDM_{\Le\Ri} - \RDM_{\Ri\Le} \right) \;.
\end{array}\end{equation}
\def\TwoReSrl{u}\def\TwoImSrl{v}%
The respective equations of motion [Eqs.\ \eqref{eq:BlochEq}] can be solved using a Laplace transformation \cite{to49}, which 
is done in \ref{sec:AppLaplace}. Under the assumption
$\GammaSo=\GammaDr=\Gamma$ one finds for the total occupation
\begin{subequations}\label{eq:DQDManPop}
\begin{eqnarray}
  s(t) &=& \e^{-(\gamma + 2 \Gamma) t} s(0) + \frac{2\Gamma}{\gamma + 2 \Gamma}
          \left[ 1 {-} \e^{-(\gamma + 2 \Gamma) t} \right]\;.\label{eq:DQDManPopSum}
\end{eqnarray}
and for the difference
\begin{eqnarray}
  w(t) &=& \frac{2 T_{\rm{c}}\eps_1}{\Omega^2}  \e^{-\gamma t} \left[ 1-\cos(\Omega t) \right] \TwoReSrl (0) \nonumber\\
        && +\frac{2 T_{\rm{c}}}{\Omega} \e^{-\gamma t} \sin(\Omega t) \TwoImSrl (0)
          +\frac{\eps_1^2 + 4 T_{\rm{c}}^2 \cos(\Omega t) }{\Omega^2} \e^{-\gamma t} w(0)\;.
\label{eq:DQDManPopDiff}
\end{eqnarray}
\end{subequations}
Together, these two expressions allow for a calculation of the occupations in the DQD.
Moreover, the current through the right barrier is given by
\begin{equation}\label{eq:DQDCurrentPhases}
J_\Dr(t)= \left\{
  \begin{array}{ll}
  \Jstat &\qquad {\rm for}\quad t\leq 0\;,\\
  e\,\gamma\,\RDM_{\Ri\Ri}(t) - e\,\Gamma \left[ 1 {-} s(t)\right] & \qquad{\rm for}\quad 0<t\leq \tp\;,\\
  e\,\Gamma\,\RDM_{\Ri\Ri}(t)  &\qquad{\rm for}\quad \tp<t\;.  
  \end{array}\right.
\end{equation}
Hereby, the stationary current is $\Jstat = e\,\Gamma\,\RDM_{\Ri\Ri}(0)=e\,\Gamma\,\RDM_{\Ri\Ri}(\infty)$. The 
typical time dependence of $J_\Dr(t)$ for $\gamma=0$ is shown in Fig.\,\ref{fig:DQDPumpProbe}e.

Thus, we are ready to get the number of pulse-induced tunneling electrons according to Eq.\,\eqref{eq:DQDNpTp}. 
This total number has contributions from the manipulation and measurement phase, i.\,e.
$\Np(\eps,\tp)=\Nmanip(\eps,\tp)+\Nmeas(\eps,\tp)$, which are according to Eq.\,\eref{eq:DQDCurrentPhases}
\begin{subequations}\label{eq:DQDTunneledElectrons}
\begin{eqnarray}
  \Nmanip(\eps,\tp)  &=& \int_0^{\tp} dt
              \left\{\gamma \RDM_{\Ri\Ri}(t) - \Gamma \left[ 1 -
      s(t)\right] \right\}\\
  \Nmeas(\eps,\tp) &=&  \int_{\tp}^\infty dt 
  \left[ \Gamma\RDM_{\Ri\Ri}(t)-\Jstat/e\right]\,.
\end{eqnarray}
\end{subequations}
The first contribution can explicitly be obtained from  Eq.\,\eqref{eq:DQDManPop}. For the case
$\gamma=0$ and $s(0)\approx 1$, one finds $s(t)\approx 1$
and consequently
\begin{equation}\label{eq:Np1Approx}
  \Nmanip (\eps,\tp) \approx 0\;.
\end{equation}
For the second contribution we use again a Laplace transform, cf.\ \ref{sec:AppLaplace}. 
The result of this procedure is the following expression for the number of
pulse-induced electrons in the measurement phase,
\begin{eqnarray}\label{eq:Curr2ndContExp}
  \Nmeas(\eps,\tp) &=&  \frac{1}{2}\left[\TwoReSrl^{2}(0)
  - \TwoImSrl^{2}(0)\right]
  - 2 \RDM_{\Le\Le}(0) \RDM_{\Ri\Ri}(0)  
  \\ &&
  - \frac{1}{2}\left[\TwoReSrl(0)\TwoReSrl(\tp)
   -\TwoImSrl(0) \TwoImSrl(\tp)\right]   
  %\nonumber\\ &&
  + \RDM_{\Ri\Ri}(0) \RDM_{\Le\Le}(\tp)
  + \RDM_{\Le\Le}(0) \RDM_{\Ri\Ri}(\tp)\,.
  \nonumber
\end{eqnarray}
Obviously, the value of $\Nmeas$ depends in a non-trivial way on {\it all} 
elements of the reduced density matrix. For weak inter-dot couplings, $T_{\rm c} \ll \Gamma$,
we can use Eq.\,\eqref{eq:DQDApprox} for the initial density matrix.
As in Eq.\,\eqref{eq:DQDApprox} we keep only terms linear in $T_{\rm c}/\Gamma$.
Thereby we obtain, together with Eq.\,\eref{eq:Np1Approx}, a compact expression for the
number of pulse-induced electrons
\begin{eqnarray}\label{eq:Np2Approx}
  \Np(\eps,\tp) 
    &\approx& \RDM_{\Ri\Ri}(\tp)
          - \frac{4 T_{\rm{c}}}{\Gamma^2 + 4 \eps^2_0} 
          \Bigl[ \eps_0 \,\TwoReSrl(\tp)
            + \frac{\Gamma}{2} \,\TwoImSrl(\tp) \Bigr]\;.
\end{eqnarray}
This is a central result of this work as it shows quantitatively the differences between $\Np$ and $\RDM_{\Ri\Ri}$
and, as we discuss in the following, explains the main features seen in the experiment. 

\begin{figure}[!t]
  \centering
  \includegraphics[width=.7\textwidth]{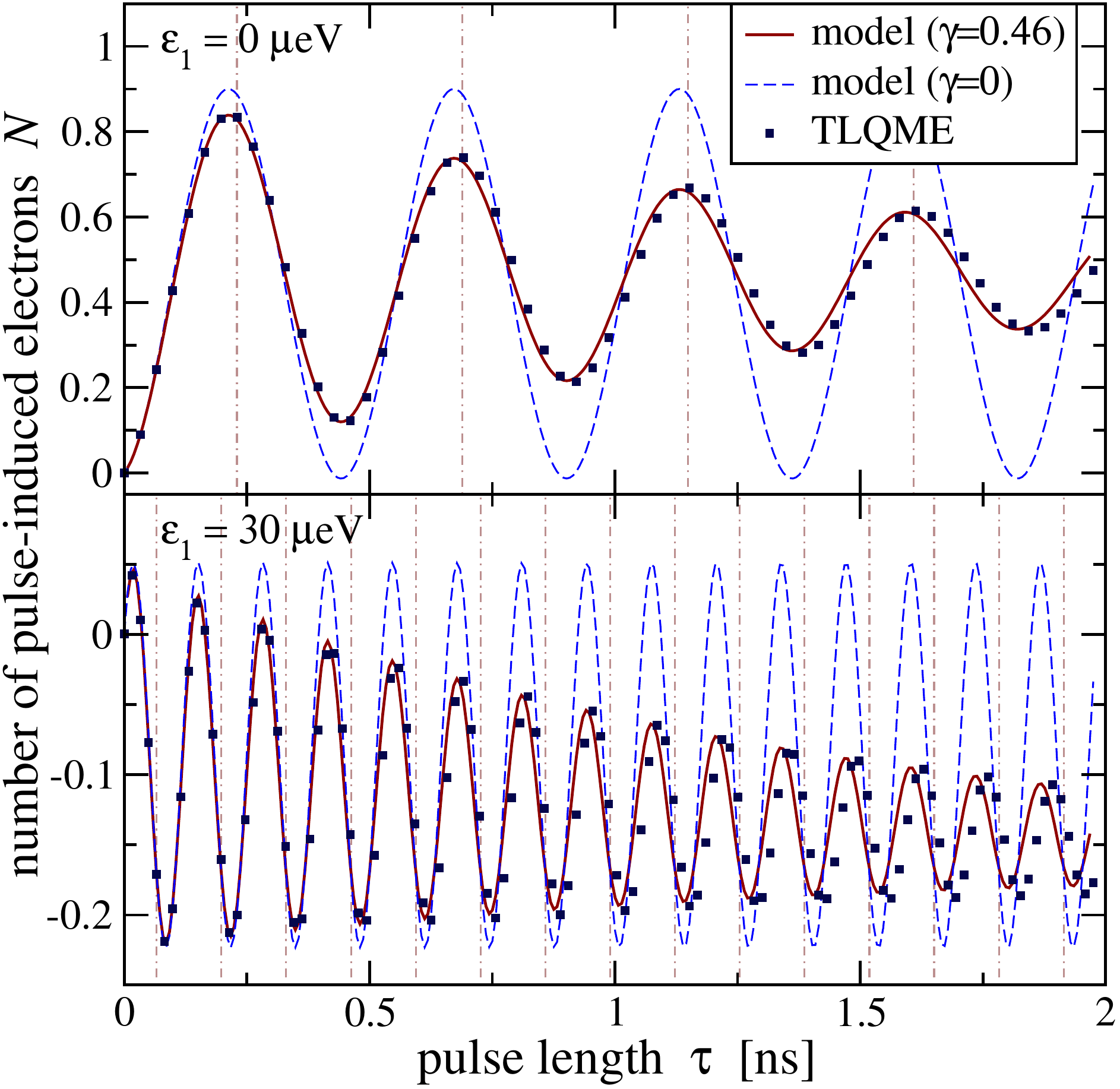}
  \caption{Number of pulse-induced electrons $\Np$ as a function of
    pulse length $\tp$. The numerically obtained results (squares) from 
    Fig.\,\ref{fig:DQDNumPopulationNp2} are shown together with the analytical expression
     [Eq.\,\eqref{eq:DQDTunneledElectrons}] of the LBL model for two damping coefficients $\gamma$ (lines).}
  \label{fig:DQDCmpPopulationNp}
\end{figure}

Firstly, Eq.\ \eqref{eq:Np2Approx} implies that for
large energy differences, $\eps_0 \gg T_{\rm{c}}$,
the number of pulse-induced
tunneling electrons indeed corresponds to the occupation of the right dot at the
end of the pulse. Therefore, it is a good measure for the occupation only for {\it non-resonant initialization} as already seen in Fig.\,\ref{fig:DQDNumPopulationNp2}.
In this case the second term in Eq.\,\eqref{eq:Np2Approx} becomes
very small. However, if the energy difference $\eps_0$, which is relevant in the 
initialization and in the measurement phase, vanishes, this term cannot be neglected. The corrections may lead
to $\Np$ taking negative values, which is shown in Fig.\,\ref{fig:DQDCmpPopulationNp}.
Secondly, for energy differences $\eps_1$ close to zero one
finds
$\Np(\eps,\tp)\approx \RDM_{\Ri\Ri}(\tp) = \frac{1}{2} [ s(\tp) - w(\tp) ] \approx \frac{1}{2} [ 1 - w(\tp) ]$. In this case,
$\Np$ is mainly determined by the population difference $w(\tp)$, which is explicitly given by Eq.\,\eqref{eq:DQDManPopDiff}.
An important consequence of this dependence is the occurrence of
an asymmetric behavior of $w(t)$ as a function of $\eps_1$ for a finite real part $u(0)$ of the  
initial coherences. For example, considering times $t_n=(2 n {+} 1)\pi/\Omega$ with $n=0,1,\ldots$
and assuming for the moment $\gamma=0$, Eq.\,\eqref{eq:DQDManPopDiff} yields 
\begin{equation}\label{eq:DQDPopAsym}
  w\left(t_n\right) = \frac{w(0)}{\Omega^2} \eps_1^2 
  +\frac{ 4 T_{\rm c} \TwoReSrl (0)}{\Omega^2}  \eps_1
  -\frac{4 T_{\rm c}^2  w(0)}{\Omega^2}\;.
\end{equation}
Obviously, this expression is only symmetric with respect to the energy difference
$\eps_1$ if $\TwoReSrl(0)=0$. In general, one finds $w\left(t_n; \eps_1\right) \neq w\left(t_n; -\eps_1\right)$.
In summary, the analysis within the described model provides a good explanation for the main
features seen in the numerical results. These features result from an additional dependence of $\Np$
on all the matrix elements of the reduced density matrix.

Finally, we will briefly discuss damping of the oscillations. 
To this end, the analytical result given by Eq.\,\eqref{eq:DQDTunneledElectrons} and the numerically obtained results from 
Fig.\,\ref{fig:DQDNumPopulationNp2} are shown together in Fig.\,\ref{fig:DQDCmpPopulationNp}. Without additional damping processes, i.\,e.\ $\gamma=0$ in Eqs.\,\eref{eq:BlochEq}, one finds for both energy differences undamped oscillations (dashed lines in Fig.\,\ref{fig:DQDCmpPopulationNp}).
Hereby, the frequency and the positions of the maxima are in good agreement
with the numerical results. For $\eps_1=30\,{\rm \mu eV}$ the
number of pulse-induced electrons, $\Np$, takes negative values. An almost
perfect agreement with the numerical results can be achieved by introducing
decoherence into the manipulation phase ($\gamma > 0$). Taking $\gamma=0.46\,{\rm \mu eV}$
(full lines in Fig.\,\ref{fig:DQDCmpPopulationNp}) yields a very good description of the pulse-length dependence of
$\Np$. Obviously, decoherence is a necessary ingredient to obtain a 
consistent picture. In the numerical calculations, the decoherence originates from
the finite source-drain voltage and the broadening of the energy levels. This
leads to a non-vanishing probability of the electron leaving the DQD \cite{vomu+05}.
In the experiment other sources of decoherence exist and these will typically dominate
the damping of coherent effects. The most important processes in this regard are
interaction of electrons with phonons, background-charge fluctuations and 
cotunneling \cite{hafu+03}. From the energy-difference dependence of the
damping rate, which is extracted from $\Np(\eps,\tp)$, one may infer about the
nature of the relevant decoherence processes \cite{hafu+03,fuha+06}.

%%%%%%%%%%%%%%%%%%%%%%%%%%%%%%%%%%%%%%%%%%%%%%%%%%%%%%%%%%%%%%%%%%%%%%%%%%%%%%%
%%% CONCLUSIONS
%%%%%%%%%%%%%%%%%%%%%%%%%%%%%%%%%%%%%%%%%%%%%%%%%%%%%%%%%%%%%%%%%%%%%%%%%%%%%%%
\section{Conclusions}\label{sec:concl}
In summary, we have theoretically investigated a pump-probe scheme realized
in a recent experiment on the coherent manipulation of charge qubits in double quantum
dots \cite{hafu+03}. To this end we have numerically simulated the pump-probe 
scheme using a time-local quantum master equation. 
The equations are solved by the auxiliary-mode expansion
technique described in \ref{sec:AuxModeProp}, which, in general, provides a flexible and viable method to study time-resolved electron transport. The numerical results
for the number of pulse-induced electrons $\Np(\eps,\tp)$
show good qualitative agreement with the experimental results.
In particular, the main feature seen in the experiment, i.\,e.\ clear oscillations
of $\Np$ as a function of pulse length $\tp$, are also observed. Moreover, two other
characteristics of the experimental results are seen in the numerical data: the asymmetry
of $\Np$ around $\eps_1 = 0$ and the occurrence of
negative values. To adress these, so far unexplained, features we considered
the Markovian limit of the QME in the respective pump-probe stages.
The resulting equations
can be solved analytically and lead to the main result of the article, namely the expression for the transferred electrons $\Np$. 
It turns out, that the value of $\Np$ depends in a non-trivial way on all elements of the reduced density matrix. 
Only for large initial energy offsets, $\eps_0 \gg T_{\rm c}$, the number of pulse-induced
tunneling electrons corresponds to the occupation of the right dot at the
end of the pulse. For small values of $\eps_0$, larger differences between the occupation and the number of pulse-induced electrons are expected, which explains the occurrence of negative values. Further analysis shows that the asymmetry can be attributed to unavoidable initial coherences resulting from the initialization stage. 
Thus, the presented findings strengthen the common interpretation of
the observed features in terms of coherent manipulation.

The expression for $\Np(\eps,\tp)$ may readily be tested against the experimental results. 
An interesting question for further studies would be the analysis of the decoherence rate as a function of energy difference. To this end, one could extent the numerical description by including phonons \cite{brkr99,lisc+08} or background-charge fluctuations. The decoherence rate may then be 
extracted from the numerical data using the analytical expression for the pulse-induced tunneling electrons from the LBL model.

%%%%%%%%%%%%%%%%%%%%%%%%%%%%%%%%%%%%%%%%%%%%%%%%%%%%%%%%%%%%%%%%%%%%%%%%%%%%%%%
%%% APPENDIX   
%%%%%%%%%%%%%%%%%%%%%%%%%%%%%%%%%%%%%%%%%%%%%%%%%%%%%%%%%%%%%%%%%%%%%%%%%%%%%%%
\appendix
\section[\mbox{}\hspace{4.5em}Auxiliary Mode Propagation]{Auxiliary Mode Propagation}\label{sec:AuxModeProp}
In order to perform the energy integration in Eqs.\ \eqref{eq:QMEDefCorrFunc} we
expand the Fermi function $f(\varepsilon)$ as a finite sum over simple poles
\begin{equation}
 f(\varepsilon)
  \approx \frac{1}{2} - \frac{1}{\beta}\sum_{p=1}^{N_\mathrm{F}}\left(
    \frac{1}{\varepsilon{-}\chi_{p}^+}
    +\frac{1}{\varepsilon{-}\chi_{p}^-}\right)\,,
  \label{eq:ExpFermiFun}
\end{equation}
with $\chi_{p}^\pm = \mu{\pm}x_p/\beta$ and $\Im x_p >0$. 
Here, $\mu$ is the chemical potential and $\beta$ the inverse electron temperature.
Instead of using the Matsubara expansion \cite{ma90}, with poles $x_p=\imath\pi(2p{-}1)$, we use a partial fraction decomposition of the 
Fermi function \cite{crsa09}, which converges much faster than the standard Matsubara expansion. In this case the poles $x_p=\pm 2\sqrt{z_p}$ are given by the eigenvalues $z_p$ of an ${N_\mathrm{F}}{\times}{N_\mathrm{F}}$ matrix \cite{crsa09}.
The poles are arranged such that all poles $\chi_p^+$ ($\chi_p^-$) are in the upper (lower)
complex plane.

Employing the expansion given by Eq.\ \eqref{eq:ExpFermiFun}, one can evaluate the energy 
integrals necessary to compute the reservoir correlation functions given by Eqs.\ \eqref{eq:QMEDefCorrFunc} by contour integration.
Thereby, the integrals in Eqs.\ \eqref{eq:QMEDefCorrFunc} become (finite) sums of the 
residues. One gets for $t>t'$
\begin{subequations}\label{eq:CorrFunExp}\begin{eqnarray}
  C^{(x y)}_{\alpha m}(t,t')
  &=&  \frac{\imath}{2} \Gamma_{\alpha m m} \delta(t-t') 
  		 - \sum_{p=1}^{N_\mathrm{F}} C^{(x y)}_{\alpha m; p}(t,t')\\
  C^{(x y)}_{\alpha m; p}(t,t') &=&
  \frac{\imath}{\beta} \Gamma_{\alpha m m}
      \e^{\imath x \int^t_{t'} dt'' \chi^x_{\alpha p} (t'')}\;,
      \label{eq:PartSelfEnerg}
\end{eqnarray}
with the auxiliary modes for reservoir $\alpha$ given by
\begin{equation}
\chi^{\pm}_{\alpha p} (t) = [\mu_{\alpha}+ \Delta_\alpha (t)] \pm x_p/\beta \pm \imath \Gamma_{\alpha m m}/2.
\end{equation}\end{subequations}
Hereby $\Delta_\alpha (t)$ is due to the time-dependent single-particle energies 
$\varepsilon_{\alpha k}(t)$ of the reservoir Hamiltonian [Eq.\,(\ref{eq:QDSResHamilOp})]. The last term containing
$\Gamma_{\alpha m m}$ leads to the correct broadening of the energy levels as discussed in Sec.\ \ref{sec:TLQME}.

With the expansion \eqref{eq:CorrFunExp} one can write the the 
auxiliary operators [Eq.\ \eqref{eq:QMEDefLAuxOp}] in terms of the partial
correlation functions. One readily obtains
\begin{equation}\label{eq:AuxOpAsSum}
  \Lambda_{\alpha m}^{(x)}(t) 
  =  \frac{1}{4} \Gamma_{\alpha m m} S^{(x)}_m
  -\sum_p \Lambda^{(x)}_{\alpha m; p}(t)
\end{equation}
with the ``partial'' auxiliary operators
\begin{equation}
 \Lambda^{(x)}_{\alpha m; p}(t):= 
 \sum\limits_{y=+,-} \;
 \int\limits_{t_0}^{t} dt' C_{\alpha m; p}^{(x y)}(t,t') \;U^\dagger_\mathrm{S}(t,t') S_m^{(y)} U_\mathrm{S}(t,t')\;.
\end{equation}
An analogous expression holds for $\tilde{\Lambda}^{(x)}_{\alpha m}$.
The partial auxiliary operators $\Lambda^{(1)}_{\alpha m; p}$ contain instead of the full correlation
function just an exponential factor (see Eq.\ \eqref{eq:PartSelfEnerg}) and the corresponding
equation of motion,
\begin{eqnarray}\label{eq:AuxOpEOM}
  \frac{\partial}{\partial t}\Lambda^{(x)}_{\alpha m; p}(t) 
  &=& \frac{\imath}{\beta} \Gamma_{\alpha m m} S_m^{(-x)}
  - \imath \left[ H_\mathrm{S}(t), \Lambda^{(x)}_{\alpha m; p}(t) \right]_- 
  \nonumber\\ &&
  +\imath x \chi^x_{\alpha p}(t) \Lambda^{(x)}_{\alpha m; p}(t)\;,
\end{eqnarray}
is easily propagated with the initial values $\Lambda^{(x)}_{\alpha m; p}(t_0)=0$.

For the numerical calculation in Sec.\ \ref{sec:Results} we represent all operators in the basis $\{\ket{0}, \ket{\Le},\ket{\Ri}\}$. This renders the equations of motion, Eq.\ \eqref{eq:QMETCL} and Eqs.\ \eqref{eq:AuxOpEOM}, to become matrix
equations. However, note that these $\sim 2N_\mathrm{F}$ equations of motion (we use $N_{\rm F} {=} 80$) are not coupled. Moreover, the density matrix $\RDM$ does not enter Eqs.\ \eqref{eq:AuxOpEOM}, which makes the propagation scheme very efficient. In order to propagate the equations
we use a fourth order Runge-Kutta scheme \cite{prfl+92} with a constant time-step $\delta t = 0.0002\,/{\rm \mu eV}$.

\section[\mbox{}\hspace{4.5em}Analytical Solution of the LBL Model]{Analytical Solution of the LBL Model}\label{sec:AppLaplace}

We summarize the solutions of the density matrix equations given in Sec.\,\ref{sec:Markovian}
and used in Sec.\,\ref{sec:model}.
They are most easily obtained using Laplace transforms  $\tilde{x}(s)=\int_0^\infty dt\:
\e^{-s t}x(t)$.

\subsection*{Manipulation Phase}

Assuming $\GammaSo=\GammaDr=\Gamma$
renders the equation for the total occupation $s$ independent from the
others, $\dot{s}(t) = - [\gamma {+} 2\Gamma] s(t) + 2 \Gamma$,
with the solution given by Eq.\ \eqref{eq:DQDManPopSum}.
The other components $w(t)$, $u(t)$, and $v(t)$ are given by coupled differential equations which
are solved by applying a Laplace transform. 
The resulting linear system reads
\begin{equation}
  \left(\begin{array}{ccc}
      s {+} \gamma & 0 & {-}2T_{\rm{c}} \\
      0& s {+} \gamma & \eps_1  \\
      2T_{\rm{c}} & {-}\eps_1 & s {+} \gamma
    \end{array}\right)
  \left(\begin{array}{c} \tilde{w}(s) \\ 
      \tilde{u}(s) \\ \tilde{v}(s) \end{array}\right)=
  \left(\begin{array}{c} w(0) \\ u(0) \\ v(0) \end{array}\right)\;.
\end{equation}
By means of the solutions  $\tilde{w}(s)$, $\tilde{u}(s)$, and $\tilde{v}(s)$ one gets
with an inverse Laplace transform the final expressions
\begin{subequations}
\mathindent=0pt
\begin{eqnarray}\label{eq:QUBITManipSol}
  w(t) &=& \e^{-\gamma t}\left[\frac{4 T^2_{\rm{c}} \cos(\Omega t) {+} \eps^2_1  }{\Omega^2}  w(0) 
        + \frac{2 T_{\rm{c}} \eps_1 \left[ 1 {-} \cos(\Omega t) \right]}{\Omega^2} u(0)	+ \frac{2 T_{\rm{c}} \sin(\Omega t) }{\Omega} v(0)\right]\,, \\
  u(t) &=& \e^{-\gamma t}\left[\frac{2 T_{\rm{c}} \eps_1 \left[ 1 {-} \cos(\Omega t) \right]}{\Omega^2} w(0)
        + \frac{4 T^2_{\rm{c}} {+} \eps^2_1 \cos(\Omega t)}{\Omega^2} u(0)        - \frac{\eps_1 \sin(\Omega t) }{\Omega} v(0)\right]\,, \\
  v(t) &=& \e^{-\gamma t}\left[- \frac{2 T_{\rm{c}} \sin(\Omega t)}{\Omega} w(0)
        + \frac{\eps_1 \sin(\Omega t) }{\Omega} u(0) 
        + \cos(\Omega t) v(0)\right]\,,
\end{eqnarray}
\end{subequations}
where we have used the Rabi frequency $\Omega$ from Eq.\,\eref{eq:rabifreq}.

\subsection*{Measurement Phase}

The Laplace transformation of Eqs.\ \eqref{eq:DQDNazarovDynEq} leads to the
following linear system
{\mathindent=20pt
\begin{equation}\label{eq:QUBITNazarovDynEqLT}
  \left(\begin{array}{cccc}
    s {+} \Gamma & \Gamma& {-}\imath T_{\rm{c}}& \imath T_{\rm{c}} \\
    0& s {+} \Gamma & \imath T_{\rm{c}}& {-}\imath T_{\rm{c}} \\
    {-}\imath T_{\rm{c}}& \imath T_{\rm{c}}& s {-}\imath \eps_0 {+} \frac{\Gamma}{2} & 0 \\
    \imath T_{\rm{c}}& {-}\imath T_{\rm{c}}& 0 & s {+} \imath \eps_0 {+} \frac{\Gamma}{2}
  \end{array}\right)
  \left(\begin{array}{c}
    \tilde{\RDM}_{\Le\Le}(s) \\
    \tilde{\RDM}_{\Ri\Ri}(s) \\
    \tilde{\RDM}_{\Le\Ri}(s) \\
    \tilde{\RDM}_{\Ri\Le}(s)
  \end{array}\right)=
  \left(\begin{array}{c}
    \RDM_{\Le\Le}(\tp) {+} \frac{\Gamma}{s} \\
    \RDM_{\Ri\Ri}(\tp) \\
    \RDM_{\Le\Ri}(\tp) \\
    \RDM_{\Ri\Le}(\tp)
  \end{array}\right)\;.
\end{equation}}%
\noindent
This system may be solved for the components $\tilde{\RDM}_{\Le\Le}(s), \tilde{\RDM}_{\Ri\Ri}(s), \tilde{\RDM}_{\Le\Ri}(s)$ and $\tilde{\RDM}_{\Ri\Le}(s)$. In order
to calculate the number of pulse-induced tunneling electrons [Eq.\ \eqref{eq:Curr2ndContExp}], one just needs $\tilde{\RDM}_{\Ri\Ri}(s)$. The stationary
value of $\RDM_{\Ri\Ri}$ for $t\to\infty$ can be calculated through the following
limit,
\begin{equation}
\lim_{s\to 0} s\, \tilde{\RDM}_{\Ri\Ri}(s) = \frac{4 T^2_{\rm{c}} }{12T^2_{\rm{c}} {+} \Gamma^2 {+} 4 \eps^2_0} = \RDM_{\Ri\Ri}(\infty)\;.
\end{equation}
This expression is identical to Eq.\ \eqref{eq:DQDInitValues}. The integrated current 
is given by
{\mathindent=20pt
\begin{eqnarray}
\lim_{s\to 0} \Gamma \left[ \tilde{\RDM}_{\Ri\Ri}(s) - \RDM_{\Ri\Ri}(\infty)/s \right] &=& 
 \frac{4}{12T^2_{\rm{c}} {+} \Gamma^2 {+} 4 \eps^2_0}\bigg[
 T^2_{\rm{c}} \RDM_{\Le\Le}(\tp) % \nonumber\\ &&
+   \left[T^2_{\rm{c}}{+} (\Gamma/2)^2 {+} \eps^2_0\right] \RDM_{\Ri\Ri}(\tp)
\nonumber\\ &&\qquad
- 2T_{\rm{c}}\left[\eps_0 \,\Re\RDM_{\Le\Ri}(\tp) + (\Gamma/2)\,\Im\RDM_{\Le\Ri}(\tp) \right]
\nonumber\\ &&\qquad
- 4 T^2_{\rm{c}} \left[ 2T^2_{\rm{c}} {+} \Gamma^2 \right]/\left[12T^2_{\rm{c}} {+} \Gamma^2 {+} 4 \eps^2_0\right]\bigg]\;.
\end{eqnarray}}%
Using the expressions for the stationary density matrix [Eqs.\,\eqref{eq:DQDInitValues}]
finally yields Eq.\,\eqref{eq:Curr2ndContExp}.

%%%%%%%%%%%%%%%%%%%%%%%%%%%%%%%%%%%%%%%%%%%%%%%%%%%%%%%%%%%%%%%%%%%%%%%%%%%%%%%
%%% REFERENCES
%%%%%%%%%%%%%%%%%%%%%%%%%%%%%%%%%%%%%%%%%%%%%%%%%%%%%%%%%%%%%%%%%%%%%%%%%%%%%%%
%\clearpage


\section*{References}
\begin{thebibliography}{10}

\bibitem{ze88}
A.~H. Zewail, Science {\bf 242},  1645  (1988).

\bibitem{sm99}
D.~L. Smith, Eng. Sci. {\bf 62},  7  (1999).

\bibitem{fuha+06}
T. Fujisawa, T. Hayashi, and S. Sasaki, Rep. Prog. Phys. {\bf 69},  759
  (2006).

\bibitem{hafu+03}
T. Hayashi {\it et~al.},
%, T. Fujisawa, H.~D. Cheong, Y.~H. Jeong, and Y. Hirayama, 
	Phys. Rev. Lett. {\bf 91},  226804  (2003).

\bibitem{pejo+05}
J.~R. Petta {\it et~al.}, Science {\bf 309},  2180  (2005).

\bibitem{kobu+06}
F.~H.~L. Koppens {\it et~al.}, Nature {\bf 442},  766  (2006).

\bibitem{zhma+05}
Y. Zhu, J. Maciejko, T. Ji, H. Guo, and J. Wang, Phys. Rev. B {\bf 71},  075317
   (2005).

\bibitem{kust+05}
S. Kurth  {\it et~al.},
%, G. Stefanucci, C.-O. Almbladh, A. Rubio, and E.~K.~U. Gross, 
	Phys. Rev. B {\bf 72},  035308  (2005).

\bibitem{hohe+06}
D. Hou, Y. He, X. Liu, J. Kang, J. Chen, and R. Han, Physica E {\bf 31},  191
  (2006).

\bibitem{mogu+07}
V. Moldoveanu, V. Gudmundsson, and A. Manolescu, Phys. Rev. B {\bf 76},  085330
   (2007).

\bibitem{stpe+08}
G. Stefanucci, E. Perfetto, and M. Cini, Phys. Rev. B {\bf 78},  075425
  (2008).

\bibitem{crsa09a}
A. Croy and U. Saalmann, Phys. Rev. B {\bf 80},  245311  (2009).

\bibitem{wesc+06}
S. Welack, M. Schreiber, and U. Kleinekath\"ofer, J. Chem. Phys. {\bf 124},
  044712  (2006).

\bibitem{jizh+08}
J. Jin, X. Zheng, and Y. Yan, J. Chem. Phys. {\bf 128},  234703  (2008).

\bibitem{nich00}
M.~A. Nielsen and I.~L. Chuang, {\em Quantum Computation and Quantum
  Information} (Cambridge University Press, 2000).

\bibitem{koma+97}
L.~P. Kouwenhoven  {\it et~al.},
%, C.~M. Marcus, P.~L. McEuen, S. Tarucha, R.~M. Westervelt, and N.~S. Wingreen,  
  in {\em Mesoscopic electron transport}, Vol.~345 of {\em
  NATO Science Series E}, edited by L.~L. Sohn, L.~P. Kouwenhoven, and G.
  Sch\"{o}n (Springer, 1997), pp.\ 105--214.

\bibitem{wide+02}
W.~G. van~der Wiel {\it et~al.},
%, F. S.~De\,Franceschi, J.~M. Elzerman, T. Fujisawa, S. Tarucha, and L.~P. Kouwenhoven,
  	Rev. Mod. Phys. {\bf 75},  1  (2002).

\bibitem{di00}
D.~P. Divincenzo, Fortschr. Phys. {\bf 48},  771  (2000).

\bibitem{grze90}
M. Gruebele and A.~H. Zewail, Physics {T}oday {\bf 43},  24  (1990).

\bibitem{cokr07}
P.~B. Corkum and F. Krausz, Nat. Phys. {\bf 3},  381  (2007).

\bibitem{kriv09}
F. Krausz and M. Ivanov, Rev. Mod. Phys. {\bf 81},  163  (2009).

\bibitem{grde92}
{\em Single charge tunneling: {C}oulomb blockade phenomena in nanostructures},
  Vol.~294 of {\em NATO Science Series B}, edited by H. Grabert and M.~H.
  Devoret (Plenum Press, 1992).

\bibitem{stna96}
T.~H. Stoof and Y.~V. Nazarov, Phys. Rev. B {\bf 53},  1050  (1996).

\bibitem{chsh79}
S. Chaturvedi and F. Shibata, Z. Phys. B {\bf 35},  297  (1979).

\bibitem{shar80}
F. Shibata and T. Arimitsu, J. Phys. Soc. Jpn. {\bf 49},  891  (1980).

\bibitem{brpe02}
H.-P. Breuer and F. Petruccione, {\em The Theory of Open Quantum Systems}
  (Oxford University Press, USA, 2002), p.\ 648.

\bibitem{brsc94}
C. Bruder and H. Schoeller, Phys. Rev. Lett. {\bf 72},  1076  (1994).

\bibitem{ovne04}
I.~V. Ovchinnikov and D. Neuhauser, J. Chem. Phys. {\bf 122},  024707  (2004).

\bibitem{lilu+05}
X.-Q. Li, J.~Y. Luo, Y.-G. Yang, P. Cui, and Y. Yan, Phys. Rev. B {\bf 71},
  205304  (2005).

\bibitem{meta99}
C. Meier and D.~J. Tannor, J. Chem. Phys. {\bf 111},  3365  (1999).

\bibitem{leko+02}
J. Lehmann, S. Kohler, P. H{\"a}nggi, and A. Nitzan, Phys. Rev. Lett. {\bf 88},
   228305  (2002).

\bibitem{wija+93}
N.~S. Wingreen, A.-P. Jauho, and Y. Meir, Phys. Rev. B {\bf 48},  8487  (1993).

\bibitem{gupr96}
S.~A. Gurvitz and Y.~S. Prager, Phys. Rev. B {\bf 53},  15932  (1996).

\bibitem{na93}
Y.~V. Nazarov, Physica B {\bf 189},  57  (1993).

\bibitem{vago+95}
N.~C. van~der Vaart {\it et~al.}, Phys. Rev. Lett. {\bf 74},  4702  (1995).

\bibitem{to49}
H.~C. Torrey, Phys. Rev. {\bf 76},  1059  (1949).

\bibitem{vomu+05}
S. Vorojtsov, E.~R. Mucciolo, and H.~U. Baranger, Phys. Rev. B {\bf 71},
  205322  (2005).

\bibitem{brkr99}
T. Brandes and B. Kramer, Phys. Rev. Lett. {\bf 83},  3021  (1999).

\bibitem{lisc+08}
G.-Q. Li, M. Schreiber, and U. Kleinekath\"{o}fer, New J. Phys. {\bf 10},
  085005  (2008).

\bibitem{ma90}
G.~D. Mahan, {\em Many-Particle Physics}, 2nd ed. (Plenum, New York, 1990).

\bibitem{crsa09}
A. Croy and U. Saalmann, Phys. Rev. B {\bf 80},  073102  (2009);
Phys. Rev. B {\bf 82},  159904  (2010).

\bibitem{prfl+92}
W.~H. Press, B.~P. Flannery, S.~A. Teukolsky, and W.~T. Vetterling, {\em
  Numerical Recipes in {C}: {T}he Art of Scientific Computing}, 2nd ed.
  (Cambridge University Press, Cambridge, 1992), p.\ 994.

\end{thebibliography}
\end{document}